# Observation of $\eta$-Al$_{41}$Sm$_5$: new evidence for structural hierarchy in Al-Sm alloys

Z. Ye[1*], F. Meng[1], F. Zhang[1], Y. Sun[1,3], L. Yang[1], S. H. Zhou[1], R. E. Napolitano[1,4], M. I. Mendelev[1], R. T. Ott[1], M. J. Kramer[1], C. Z. Wang[1], K. M. Ho[1,2,3†]

[1]Ames Laboratory, US Department of Energy, Ames, Iowa 50011, USA

[2]Department of Physics, Iowa State University, Ames, Iowa 50011, USA

[3]Hefei National Laboratory for Physical Sciences at the Microscale and Department of Physics, University of Science and Technology of China, Hefei, Anhui 230026, China

[4]Department of Materials Sci. and Eng., Iowa State University, Ames, Iowa, 50011, USA

Authors to whom correspondence should be addressed:
[*] Email address: zye@iastate.edu (Zhuo Ye)
[†] Email address: kmh@iastate.edu (Kai-Ming Ho)

Abstract

Using an effective genetic algorithm, we uncover the structure of a metastable $\eta$-$Al_{41}Sm_5$ phase that supplements its family sharing similar short-range orders. The $\eta$-phase evolves upon heating an amorphous Al-9.7at.%Sm ribbon, produced by melt-spinning. The dynamical phase selection is discussed with respect to the structural connections between the short-range packing motifs in the amorphous precursor and those observed in the selected phases. The $\eta$-phase elucidated here is one of several newly discovered large-unit-cell phases found to form during devitrification from the glass in this binary system, further illustrating the power and efficiency of our approach, the important role of structural hierarchy in phase selection, and the richness of the metastable phase landscape accessible from the glassy structure.

## 1. Introduction

Growing demand for advanced materials with enhanced functionality promotes expansion of the set of accessible structures. While stable materials have been efficiently identified and produced, meta-stable states are considered as a big challenge to be predicted and realized. Glass-forming alloys offer a rich landscape of non-equilibrium states, both crystalline and non-crystalline, and far-from equilibrium pathways to access them. The pathways can be manipulated through changing the starting points of materials, such as the processing parameters and the chemical composition of alloy. Al-Sm alloys, known as marginal glass formers, provide a prototypical model system where a rich collection of intermediate meta-stable crystalline phases can be accessed through path-dependent devitrification processing [1–4]. A fundamental scientific question is, what is the underlying physics mechanism of phase selections in this far-from-equilibrium system and how to control the pathways to access a myriad of meta-stable structures. A basic understanding of the physical principles that govern these pathways could very well enable application of the same principles to many different systems.

As reported in our recent work [5], the short-range order (SRO) that develops in an undercooled liquid and glass plays an important role in phase selection during devitrification processes. The amorphous Al-Sm alloys realized in melt-spun ribbon and magnetron sputtered thin films devitrify following completely different pathways. Constant-heating-rate (CHR) devitrification of Al-10.2at% Sm ribbon exhibits a polymorphic transformation that results in a cubic $\varepsilon$-$Al_{60}Sm_{11}$ phase [6] with a lattice parameter of ~14Å, space group $Im\bar{3}m$ (No. 229) and with 6 unique Wyckoff positions. The thin film of the same chemical composition develops compositional inhomogeneities before the formation of fcc-Al and a hexagonal $\theta$-$Al_5Sm$ phase [7] with space group $P6_322$ (No. 182) and with 5 Wyckoff positions. Selection is also composition dependent, and the $\theta$-phase is observed as the initial crystallized phase with fcc-Al during CHR devitrification of an Al-14.1% Sm melt-spun ribbon. The $\varepsilon$-$Al_{60}Sm_{11}$ and $\theta$-$Al_5Sm$ phases share the same Sm-centered first-shell atomic packing motif, termed "3-6-6-1" in Ref. 5. The same 3-6-6-1 motif is found dominant in undercooled liquids, indicating a clear structural inheritance from the liquid to its devitrified crystalline phases.

We have recently developed an approach to solve for undetermined complex crystal structures observed in far-from equilibrium transitions [6,7]. The approach integrates lattice and space group information from X-ray diffraction (XRD) analysis with a genetic algorithm (GA) structural search. With this approach, we have successfully identified several previously unknown LUC structures, including the $\varepsilon$-$Al_{60}Sm_{11}$ phase [6] and the $\theta$-$Al_5Sm$ phase (i.e. $Al_{20}Sm_4$ in Ref. [7]). Presently, we report on the discovery and identification of a LUC (~90 atoms/cell) tetragonal structure, termed hereafter as $\eta$-$Al_{41}Sm_5$ in this work.

The unknown phase appears as a part of a polyphase assembly of metastable phases that evolve during devitrification of an amorphous Al-9.7%Sm melt spun ribbon. The existence of common Sm-centered first-shell packing motifs in each of these LUC crystal structures ($\varepsilon$, $\theta$, and $\eta$ phases) provides clear evidence for the critical role of structural hierarchy in complex phase selection in Al-Sm alloys. In this picture, specific short-range packing motifs which contribute to glassy behavior in undercooled liquids and amorphous solids also serve as precursors for particular crystalline phases that appear during initial stages of devitrification.

## 2. Experimental

2.1. Preparation of the amorphous alloy

Alloys of Al-9.7%Sm were prepared in ingots of 10 grams by simultaneously arc-melting the pure components (99.99 wt% Al and 99.9 wt% Sm) under an Ar atmosphere. The alloy ingots were re-melted under 1/3 atm argon and rapidly solidified into ribbons (0.02-0.03 mm thick) by free-jet melt-spinning onto a rotating Cu wheel (tangential speed of 30 m/s). Both high resolution transmission electron microscopy (TEM) and high energy X-ray diffraction (HEXRD) reveal the amorphous nature without any detectable crystallized phase (not shown here).

2.2. Crystallization of the amorphous alloy

Crystallization of the amorphous alloy ribbons was investigated using time-resolved synchrotron-based HEXRD (71.77 keV energy, 0.1729 Å wavelength), utilizing the 1-ID-E beamline of the Advanced Photon Source (APS) at Argonne National Laboratory. Specimens for HEXRD were prepared by cutting melt-spun ribbon into lengths of approximately 10 mm, stacking multiple segments to a thickness of ~0.5mm, and inserting into a 2mm ID thin walled $SiO_2$ capillary tube which was sealed under argon. An infrared heater was used for in-situ heating and isothermal holding. Post-devitrification samples were analyzed using TEM (Tecnai $G^2$ F20). TEM specimens were prepared using a dual-beam focused ion beam (FIB) instrument (FEI Helios NanoLab G3 UC).

Upon CHR heating, the as-quenched amorphous ribbon exhibits a multi-step devitrification pathway that is characterized by a series of metastable crystalline phases [2,8]. The $\varepsilon$-$Al_{60}Sm_{11}$ phase and a small amount of fcc-Al are the first phases to appear during devitrification. However, there are several minor XRD peaks that cannot be indexed, as marked by diamonds in Fig. 1(a), suggesting a small fraction of an unknown phase mixed with the $\varepsilon$-$Al_{60}Sm_{11}$ phase and Al-fcc. With an isothermal hold at 464 K (about 5 K lower than the onset temperature for crystallization) for 70 mins, the unknown phase grows as indicated in the enhanced peaks in XRD as shown in Fig. 1(b). More peaks of the unknown phase are also observed to appear in Fig. 1(b). Figure 2 (a)-(c) show the bright field transmission electron microscopy (BF-TEM) image, high angle annular dark field scanning transmission electron microscopy (HAADF-STEM) image and selected area diffraction (SAD) pattern, respectively. The BF-TEM image shown in Fig. 2(a) and the HAADF-STEM image in Fig. 2(b) clearly demonstrate a fully crystallized polyphase material. The atomic number contrast in the HAADF-STEM image indicates the presence of three phases, corresponding to the bright, dark and gray regions. The SAD pattern in Fig. 2(c) shows the diffraction patterns of the $\varepsilon$-phase and the unknown phase.

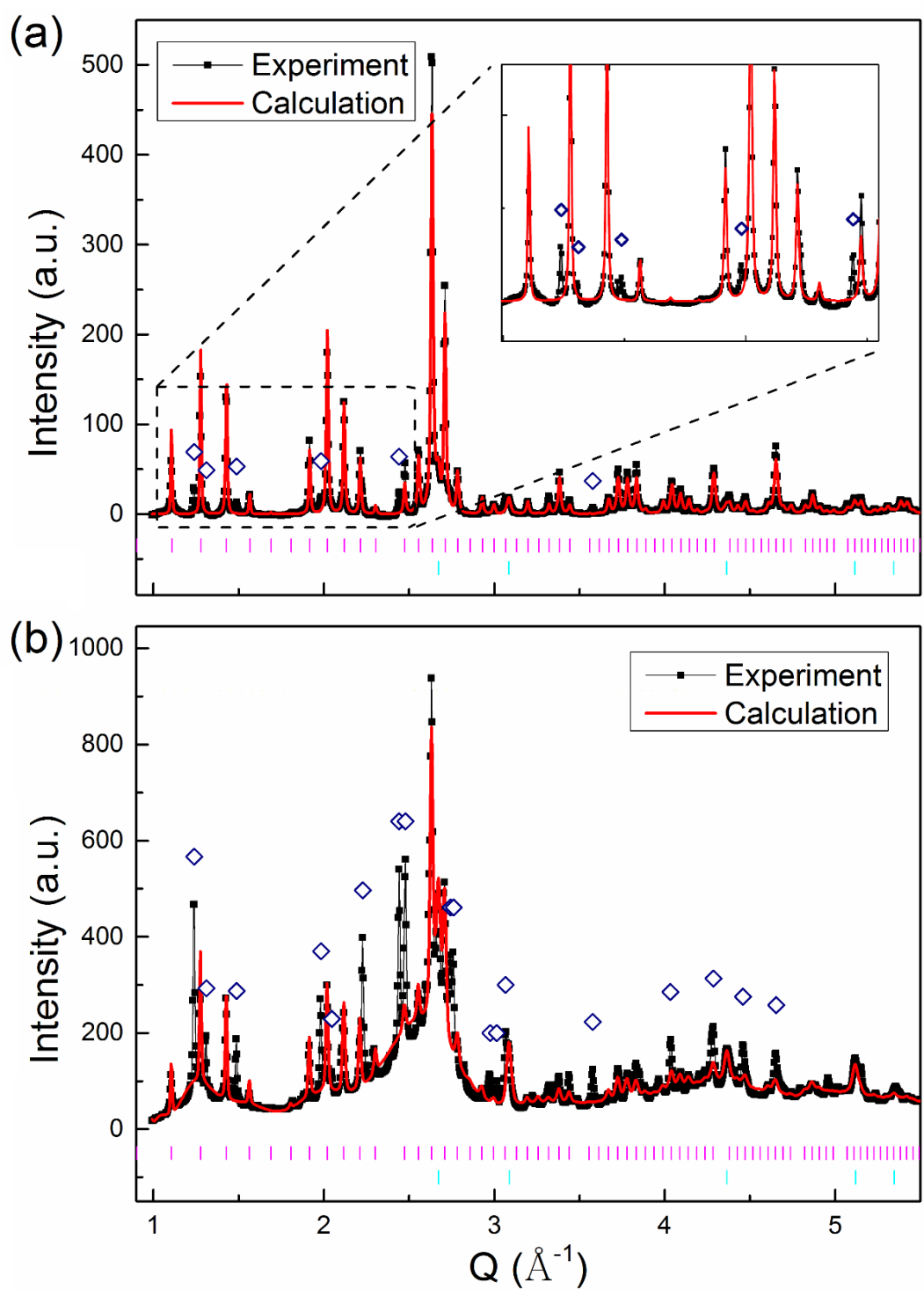

Fig. 1. XRD patterns where (a) the unknown phase begins to appear, and (b) the unknown phase grows after an isothermal hold. Inset of Fig. 1(a): Amplified view of the low $Q$ region, showing the peaks of the unknown phase. Black lines show the XRD pattern, while red lines show the Rietveld fitting of the XRD with $\varepsilon$-$Al_{60}Sm_{11}$ and fcc-Al. The vertical lines in cyan and navy show the peak positions for $\varepsilon$-$Al_{60}Sm_{11}$ and Al, respectively. The magenta diamonds indicate the diffraction peaks of the unknown phase.

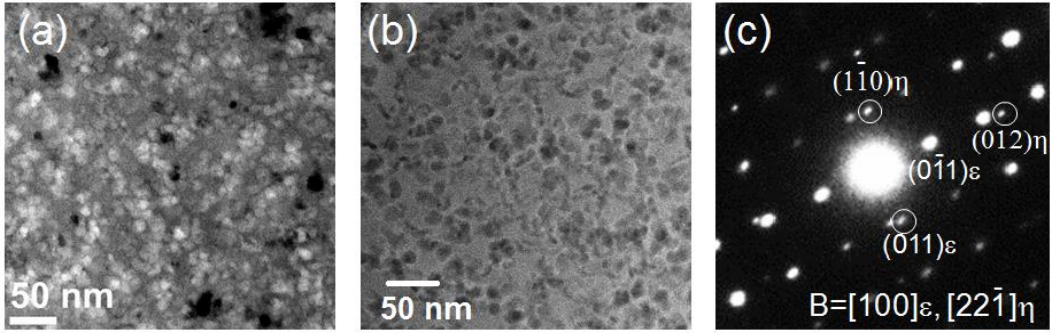

Figure 2. (a) Bright-field transmission electron microscopy and (b) high-angle annular dark-field (HAADF) scanning transmission electron microscopy images of a polyphased matrix of the metastable phases. (c) The corresponding selected area electron diffraction (SAD) pattern of the $\varepsilon$-$Al_{60}Sm_{11}$ phase and the unknown phase (designated $\eta$-$Al_{41}Sm_5$ later in this work). The zone axis is $[100]$ and $[22\bar{1}]$ for the $\varepsilon$- and the $\eta$-phase, respectively.

## 3. Results and discussion

3.1. Identification of the unknown phase

The complex structure of the unknown phase was determined by employing the approach described in Ref. [6]. Using standard space group peak-matching techniques to analyze the XRD pattern shown in Fig. 1(b), we initially identify the unit cell as body-centered tetragonal with lattice parameters, a = 13.33 Å and c = 9.59 Å. Based on an assumed density equal to that of the glass (0.051 atoms/Å$^3$), we estimate the number of atoms per unit cell to be approximately 90. Using a classical interatomic potential for computational expediency [9], we perform a GA search, seeking low-energy structures with the tetragonal unit cell and the space group of $I\,4$ and $I\,\bar{4}$. The search includes computing an XRD pattern for each structure in the GA pool using the Rietveld program RIETAN-FP [10]. A profile factor $F_{XRD}$ is calculated to assess how well the computed pattern fits the experiment measurements [6,7]. A lower $F_{XRD}$ indicates a better match with the experimental XRD pattern. After selecting a small set of candidate structures based on the $F_{XRD}$, a more accurate first-principles density function theory (DFT) energy is calculated [11] using the Vienna *ab initio* simulation package (VASP) [12] with the projector-augmented wave (PAW) pseudopotential method [13,14] and the generalized-gradient approximation (GGA) [15].

The crystal structure exhibiting the lowest formation energy in the GA pool is shown in Fig. 3(a). The phase, designated here as $\eta$, exhibits a tetragonal unit cell with space group No. 87 (I4/m) including 10 Wyckoff positions and a stoichiometry of $Al_{82}Sm_{10}$. With two formula units per unit cell, the complete designation for this phase becomes $\eta$-$Al_{41}Sm_5$. The formation energy is 0.051 eV/atom with respect to fcc-Al and $Al_3Sm$. It also has a low $F_{XRD}$, which indicates a good match with experiment XRD, as shown in Fig. 4. Fig. 4(d) shows the formation energy with respect to trigonal-Sm and fcc-Al of known stable, metastable phases, and the recently solved metastable phases $\pi$-$Al_5Sm$ [4], $\varepsilon$-$Al_{60}Sm_{11}$ [6], $\theta$-$Al_5Sm$ [7], and $\eta$-$Al_{41}Sm_5$. To distinguish the $2b$ and $8h$ Sm sites, they are marked in Fig. 3 (a) with blue and grey, respectively. We highlight here the first shell packing environments around the 2 Sm sites, as illustrated with blue and grey polyhedral. The motif around the $2b$ and $8h$ Sm site is termed as 1-6-6-6-1 and 1-5-6-5-1, respectively, based on the packing of the atoms around the Sm atom as shown in Fig. 3(b) and (c).

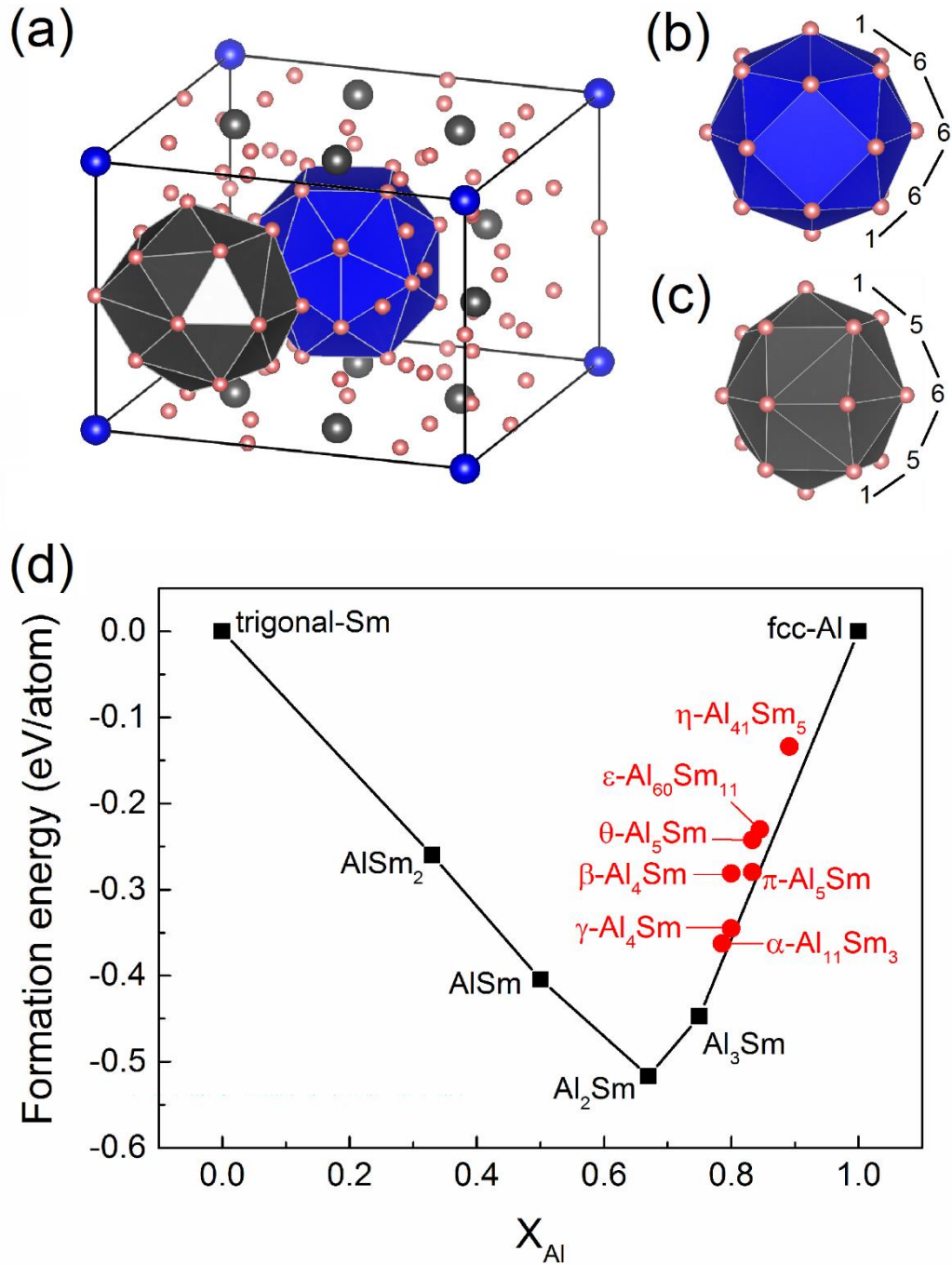

Figure 3. (a) The structure of $\eta$-Al$_{41}$Sm$_5$ showing two Sm-centered motifs: (b) the 1-6-6-6-1 motif marked in blue, and (c) the 1-5-6-5-1 motif in grey. Pink (blue/grey) represents Al (Sm) atoms. (d) The formation energy of known stable and meta-stable phases at 0 K as a function of the Al composition. The solid line connects the thermodynamically stable phases (black squares) shown in the phase diagram. The red circles are the meta-stable phases.

3.2. Rietveld refinement

    A Rietveld fitting is done to refine the lattice and atomic positions. We choose to fit the data at

464 K, and the result in Fig. 4 reveals three different phases: the cubic $\varepsilon$-$Al_{60}Sm_{11}$, fcc-Al and the tetragonal $\eta$-$Al_{41}Sm_5$, which constitute ~37.1, 28.3, and 34.6 wt. %, respectively. Table I shows the lattice parameters and atomic coordinates of the $\eta$-$Al_{41}Sm_5$ phase, given by both DFT calculations and the Rietveld analysis.

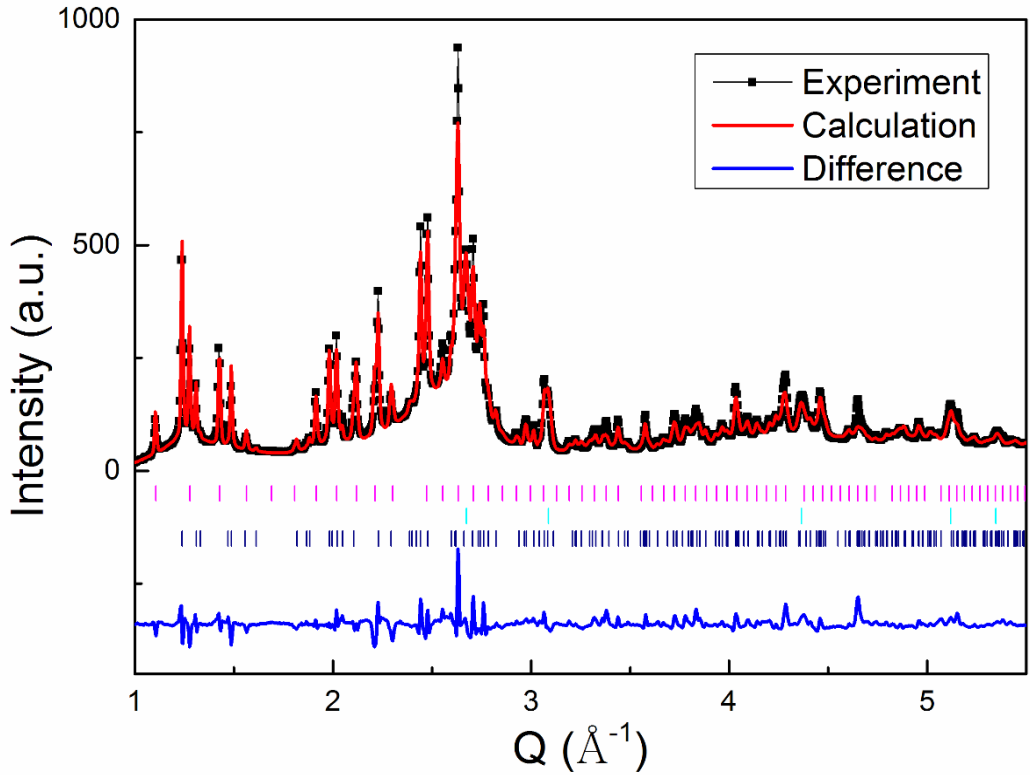

Fig. 4. Rietveld fitting of the XRD patterns for melt spun Al-9.7%Sm ribbon at 464 K, showing the crystallization products of the cubic $\varepsilon$-$Al_{60}Sm_{11}$, fcc-Al and the tetragonal $\eta$-$Al_{41}Sm_5$. The vertical lines in magenta, cyan, and navy show the diffraction peak positions for $\varepsilon$-$Al_{60}Sm_{11}$, Al and $\eta$-$Al_{41}Sm_5$, respectively. The fitting of the experimental data gives wRp = 0.0914 and Rp = 0.0647, where wRp and Rp are the weighted and unweighted profile R-factors, respectively.

| Lattice parameters (in unit of Å) | | | | |
|---|---|---|---|---|
| a = 13.284 (13.348), c = 9.568 (9.588) | | | | |
| Atomic coordinates | | | | |
| | X | Y | Z | Wyckoff |
| Al1 | 0.619 (0.603) | 0.282 (0.284) | 0 | 8h |
| Al2 | 0.097 (0.089) | 0.522 (0.515) | 0 | 8h |
| Al3 | 0.136 (0.140) | 0.074 (0.080) | 0.191 (0.180) | 16i |
| Al4 | 0.572 (0.565) | 0.136 (0.138) | 0.252 (0.236) | 16i |
| Al5 | 0.564 (0.563) | 0.725 (0.722) | 0.145 (0.148) | 16i |
| Al6 | 0.25 | 0.25 | 0.25 | 8f |
| Al7 | 0 | 0 | 0 | 2a |
| Al8 | 0.840 (0.848) | 0.282 (0.290) | 0 | 8h |
| Sm1 | 0.927 (0.925) | 0.722 (0.724) | 0 | 8h |
| Sm2 | 0 | 0 | 0.5 | 2b |

Table 1. Lattice parameters and atomic coordinates of the $\eta$-Al$_{41}$Sm$_5$ phase with the space group No. 87 (I4/m). The numbers in parentheses are given by Rietveld analysis, and the rest by DFT calculations.

3.3. Short range order in Al-Sm glass and its devitrified crystals

While lacking long-range translational symmetry, the as-quenched Al-Sm glasses, like many glasses, have clear elements of short range order (SRO). In particular, these Sm-centered packing motifs, play a crucial role in phase selection during devitrification. In simulated undercooled Al-10%Sm liquids, the "3-6-6-1" motif is the dominant Sm-centered motif [5]. Experimentally the Al-Sm glasses can be synthesized by melt spinning and magnetron sputtering. Fcc-Al and the $\theta$-Al$_5$Sm phase precipitate from the amorphous sputtered Al-10%Sm thin film [7]. The $\theta$-Al$_5$Sm structure is composed exclusively of the same "3-6-6-1" motif, indicating a well-defined structural order that transcends glass and its devitrified crystalline phase. The $\varepsilon$-Al$_{60}$Sm$_{11}$ phase [6] precipitates from the amorphous melt spun Al-10.2%Sm ribbon. The $\varepsilon$-Al$_{60}$Sm$_{11}$ phase also exhibits the same "3-6-6-1" motif, indicating a clear structural inheritance from the glass. In addition, it has another "1-6-6-6-1" motif [6]. The $\eta$-Al$_{41}$Sm$_5$ phase appears along with the $\varepsilon$-Al$_{60}$Sm$_{11}$ phase devitrified from the melt spun Al-9.7%Sm ribbon. The $\eta$-Al$_{41}$Sm$_5$ phase is composed of the same "1-6-6-6-1" as in the $\varepsilon$-Al$_{60}$Sm$_{11}$ phase and a new "1-5-6-5-1" motif as shown in Fig. 3(a)-(c). The appearance of similar cluster motifs between the $\varepsilon$-Al$_{60}$Sm$_{11}$ and the $\eta$-Al$_{41}$Sm$_5$ phases provides another solid evidence for the structural hierarchy mechanism of complex phase selections in Al-Sm alloys. The question remains, however, as to whether the $\eta$-Al$_{41}$Sm$_5$ phase is formed directly from the glass or the $\varepsilon$-Al$_{60}$Sm$_{11}$ phase is a precursor for the formation of the $\eta$-Al$_{41}$Sm$_5$ phase. This question calls for additional investigation.

3.4. Free energy calculation

At 0 K, the $\eta$-Al$_{41}$Sm$_5$ phase is 0.051 eV/atom unstable with respect to phase separation into the Al$_3$Sm phase and pure Al. To investigate the effects of finite temperatures, we calculated the Gibbs free energy energy within the quasi-harmonic approximation using the Phonopy package [16].

At a fixed volume, the Holmoltz free energy under the harmonic approximation is given by

$$F = E_0 + k_B T \sum_{nk} \ln\left[2\sinh\frac{\hbar\omega_n(k)}{2k_B T}\right], \tag{1}$$

where $E_0$ is the zero-temperature total energy from VASP calculation, and $\omega_n(k)$ is the phonon spectrum. To account for thermal expansion, the phonon spectrum is calculated at various volumes, and the Gibbs free energy $G$ is obtained by minimizing $F$ with respect to the volume [16]. In Fig.5 (a), we first show the phonon density of states of the $\eta$-$Al_{41}Sm_5$ phase at 0 K. No negative phonon modes were observed, indicating that the $\eta$-$Al_{41}Sm_5$ phase is mechanically stable. In Fig. 5 (b), we plot the formation Gibbs free energy, referenced to fcc-Al and $Al_3Sm$, as a function of temperature, where one can see that the $\eta$-$Al_{41}Sm_5$ phase remains unstable w.r.t. Al and $Al_3Sm$ ($G_{form} > 0$) for the entire temperature range of the devitrification process. However, $G_{form}$ decreases with the temperature, showing that it becomes more stable as temperature increases.

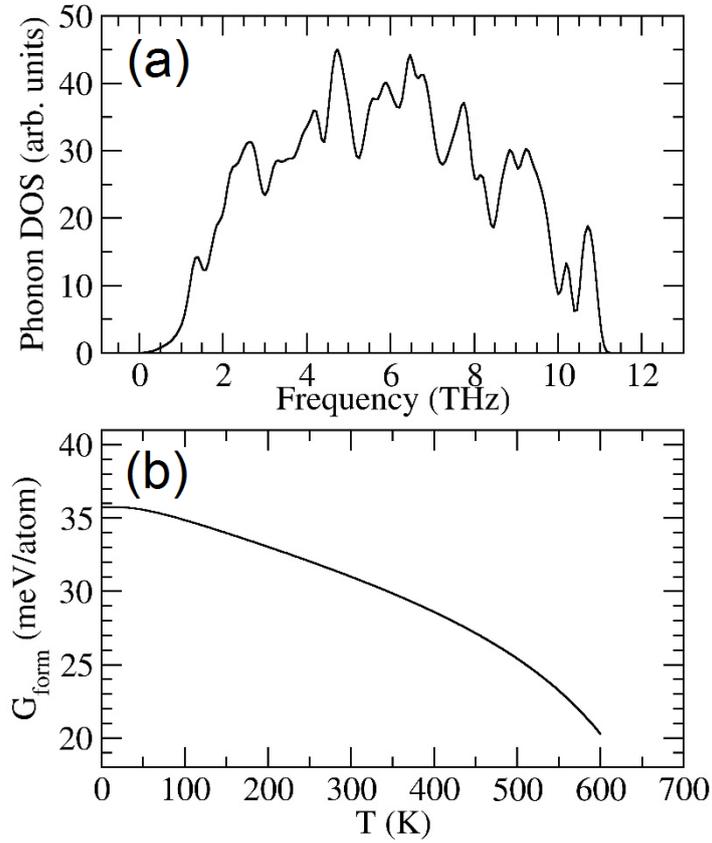

Figure 5. (a) Phonon density of states of the $\eta$-$Al_{41}Sm_5$ phase. (b) The formation Gibbs free energy as a function of the temperature referenced to Al and $Al_3Sm$.

## 4. Summary

We solve an unknown metastable phase observed during the devitrification of melt spun Al-10%Sm glasses using an efficient genetic algorithm combined with experimental diffraction data. The crystalline phase has a big tetragonal unit cell that contains 92 atoms with the stoichiometry $Al_{82}Sm_{10}$. The calculated X-ray diffraction pattern matches well with that of the experiments. $\eta$-$Al_{41}Sm_5$ is mechanically stable from phonon spectrum calculation. It is metastable with respect to phase separation into $Al_3Sm$ and pure Al at finite temperatures. Resolution of the atomic structure of these new metastable complex crystal phases lay the groundwork for further investigations to elucidate how different processing protocols can influence the selection and growth of different metastable crystal phases in the devitrification process. Examining the 3 metastable phases observed so far in devitrification experiments, we find a common picture emerging where complex metastable phases which appear have structures dominated by specific atomic clusters (motifs) centered about Sm atoms. This supports structure hierarchy picture of complex phase formations and suggests a possible physical mechanism where the low mobility of Sm atoms during devitrification process plays an important role in the selection of metastable crystal phases compatible with certain Sm-centered cluster motifs.

Acknowledgement

Work at Ames Laboratory was supported by the US Department of Energy, Basic Energy Sciences, Materials Science Division and Engineering, under Contract No. DE-AC02-07CH11358, including a grant of computer time at the National Energy Research Supercomputing Center (NERSC) in Berkeley, CA. The high-energy X-ray experiments were performed at the XOR beamline (sector 1) of the Advanced Photon Source, Argonne National Laboratory, under Grant No. DE-AC02-06CH11357. K.M.H. acknowledges support from USTC Qian-Ren B (1000-Talents Program B) fund.